\documentclass[11pt]{article}

\usepackage{amsmath,amsfonts,amssymb,latexsym}

\newtheorem{theorem}{Theorem}[section]
\newtheorem{definition}{Definition}[section]

\newtheorem{corollary}[theorem]{Corollary}

\newenvironment{proof}[1][Proof]{\textsc{#1.} }{\ \rule{0.5em}{0.5em}}
\numberwithin{equation}{section}

\begin{document}

\title{{\Huge Global Hyperbolicity and Completeness}}
\author{{\Large Yvonne Choquet-Bruhat} \\
University of Paris VI\\ Tour 22-12\\ 4 Place Jussieu, 75252
Paris\\ France\\ \texttt{email:\thinspace ycb@ccr.jussieu.fr} \and
{\Large Spiros Cotsakis} \\ \textsc{GEODYSYC}\\ Department of
Mathematics\\ University of the Aegean\\ Karlovassi 83 200, Samos,
Greece\\ \texttt{email:\thinspace skot@aegean.gr}} \maketitle

\begin{abstract}
\noindent We prove global hyperbolicity of spacetimes under generic
regularity conditions on the metric. We then show that these spacetimes are
timelike and null geodesically complete if the gradient of the lapse and the
extrinsic curvature $K$ are integrable. This last condition is required only
for the tracefree part of $K$ if the universe is expanding.
\end{abstract}

\section{Introduction}

The global existence problem in General Relativity does not reduce to a
global existence theorem for a solution of the Einstein equations with some
choice of time coordinate. The physical problem is the existence for an
infinite proper time. But the proper time depends on the observer, i.e. on
the timelike line along which it is observed. The notorious singularity
theorems consider as a singularity the timelike or null geodesic
incompleteness of a spacetime. They are established under various geometric
hypothesis. On the other hand, existence theorems for a solution of generic
Cauchy problems for the Einstein equations with an infinite proper time of
existence for a family of observers have been obtained but their timelike or
null geodesic completeness has not been proved, at least explicitly, in
dimensions greater than 1+1 (cf. \cite{ch-kl93,cb-mo01}).

We give in this paper generic conditions under which a spacetime
is globally hyperbolic and sufficient conditions under which it is
timelike and null geodesically complete.  In the special case of
an expanding universe our sufficient conditions for future
completeness can be  weakened, it  applies in particular to the
spacetimes constructed in \cite{cb-mo01}. If one would find
corresponding necessary conditions for completeness along the
lines we use to obtain sufficient conditions, they would give a
generic singularity theorem, based on analysis.

\section{Global hyperbolicity}

We consider a spacetime $(\mathcal{V},g)$ with $\mathcal{V}=\mathcal{M}%
\times \mathcal{I},\;$ $\mathcal{I}=(t_{0},\infty )$, where $\mathcal{M}$ is
a smooth manifold of dimension $n$ and $^{(n+1)}g$ a Lorentzian metric which
in the usual $n+1$ splitting reads,
\begin{equation}
^{(n+1)}g\equiv -N^{2}(\theta ^{0})^{2}+g_{ij}\;\theta ^{i}\theta ^{j},\quad
\theta ^{0}=dt,\quad \theta ^{i}\equiv dx^{i}+\beta ^{i}dt.  \label{2.1}
\end{equation}
The spacetime is time-oriented by increasing $t.$ We have chosen
$\mathcal{I} =(t_{0},\infty )$ because we had in mind the case of
an expanding universe with a singularity in the past, for instance
at $t=0<t_{0}.$ However, since $ t$ is just a coordinate, our
study could apply as well to any interval $
\mathcal{I}\subset\mathbb{R}$. To avoid irrelevant writing
complications we suppose the metric to be as smooth as necessary
and we make the following hypotheses which exclude pathologies in
the metric and its representative:

\begin{itemize}
\item  \textbf{Bounded lapse:} The lapse function $N$ is bounded below and
above by positive numbers $N_{m}$ and $N_{M}$,
\begin{equation}
0<N_{m}\leq N\leq N_{M}.  \label{h1}
\end{equation}
This hypothesis insures that the parameter $t$ measures, up to a positive
factor bounded above and below, the proper time along the normals to the
slices $\mathcal{M}_{t}\,(=\mathcal{M}\times \{t\}).$

\item  \textbf{Slice completeness:} The time dependent metric $g_{t}\equiv
g_{ij}dx^{i}dx^{j}$ is a complete Riemannian metric on $\mathcal{M}_{t}\,(=%
\mathcal{M}\times \{t\})$, uniformly bounded below for all $t\in \mathcal{I}$
by a metric $\gamma =g_{t_{0}}$. That is we suppose that there exists a
number $A>0$ such that for all tangent vectors $v$ to $\mathcal{M}$ it holds
that
\begin{equation}
A\gamma _{ij}v^{i}v^{j}\leq g_{ij}v^{i}v^{j}.  \label{h2}
\end{equation}

\item  \textbf{Uniformly bounded shift:} The $g_{t}$ norm of the shift
vector $\beta$, projection on the tangent space to $\mathcal{M}_{t}$ of the
tangent to the lines $\{x\}\times \mathcal{I}$, is uniformly bounded by a
number $B.$
\end{itemize}

Under these hypotheses we prove the following theorem.

\begin{theorem}
The spacetime $(\mathcal{V}\equiv\mathcal{M}\times\mathcal{I},\,^{(n+1)}g)$
is globally hyperbolic.
\end{theorem}

We will use Geroch's criterion of global hyperbolicity \cite{ge70},
equivalent to Leray's original definition \cite{le52}. We will show that
each $\mathcal{M}_{t},$ $t\in \mathcal{I}$, is a Cauchy surface of the
spacetime. In fact $\mathcal{M}_{t}$ is a spacelike submanifold because its
normal $n$ is timelike: If $n$ is future-directed its components in our
frame are, $n_{0}=-N,$ $n_{i}=0$.

\begin{proof}
We now prove that $\mathcal{M}_{t}$ is cut once by every inextendible causal
curve. Indeed, if $C:\mathcal{I}\rightarrow \mathcal{V}:\lambda \mapsto
C(\lambda )$ is a future-directed causal curve, its tangent is such that,
\begin{equation}
^{(n+1)}g\left( \frac{dC}{d\lambda },n\right) \equiv -N\frac{dt}{d\lambda }
<0,
\end{equation}
therefore on $C$ we have that,
\begin{equation}
\frac{dt}{d\lambda }>0,
\end{equation}
and hence $C$ can be reparametrized using $t$ and cuts each $\mathcal{M}_{t}$
at most once.

To show that a future-directed causal curve $C$ issued from a point $%
(x_{1},t_{1})$ cuts each $\mathcal{M}_{t}$ with $t>t_{1}$ we prove that any
such curve can be extended to $t=+\infty $ (an analogous reasoning will
prove that a past-directed causal curve cuts each $\mathcal{M}_{t}$ with $%
t_{0}<t<t_{1}$). Indeed, suppose the curve is defined for $t\in \lbrack
t_{1},T)$, with $T$ finite (respectively $T>t_{0})$. Consider a Cauchy
sequence of numbers $(t_{n})$ which converges to $T$ and the corresponding
points ($c_{n},t_{n})$ of the curve $C,$ where $c_{n}$ (with components $%
C^{i}(t_{n})$) are points of $M$. To show that when $(t_{n})$ is a Cauchy
sequence converging to $T$, these points converge to a limit point, denoted
by $c(T)$, we consider the distance $d$ in the complete metric space $(%
\mathcal{M},\gamma )$ and we have:
\begin{equation}
d(c_{n},c_{m})\leq \int_{t_{n}}^{t_{m}}\left( \gamma _{ij}\frac{dC^{i}}{dt}
\frac{dC^{j}}{dt}\right) ^{1/2}dt.
\end{equation}
We deduce from hypothesis (\ref{h2}) that for all $t$,
\begin{equation}
\gamma _{ij}\frac{dC^{i}}{dt}\frac{dC^{j}}{dt}\leq A^{-1}g_{ij}\frac{dC^{i}}{%
dt}\frac{dC^{j}}{dt},
\end{equation}
and from the causality of $C$ and assumption (\ref{h1}) we obtain,
\begin{equation}
\left( g_{ij}\left( \frac{dC^{i}}{dt}+\beta ^{i}\right) \left( \frac{dC^{j}}{%
dt}+\beta ^{j}\right) \right) ^{1/2}\leq N\leq N_{M}.
\end{equation}
The classical subadditivity of norms implies then that,
\begin{equation}
\left( g_{ij}\frac{dC^{i}}{dt}\frac{dC^{j}}{dt}\right) ^{1/2}\leq N+\left(
g_{ij}\beta ^{i}\beta ^{j}\right) ^{1/2}\leq N_{M}+B.
\end{equation}
Assembling these results we find that,
\begin{equation}
d(c_{n},c_{m})\leq A^{-1}(N_{M}+B)(t_{m}-t_{n}).
\end{equation}
This inequality shows that if $(t_{n})$ is a Cauchy sequence of numbers
converging to $T,$ the sequence $(c_{n})$ is a Cauchy sequence in the
complete Riemannian manifold $(\mathcal{M},\gamma )$, hence admits a limit
point which we call $c(T).$ The timelike curve on $[t_{1},T)$ is therefore
extendible.
\end{proof}

\section{Completeness}

We have seen that all future-directed timelike curves issued at time $t_{1}$
can be represented by a mapping $C:[t_{1},+\infty )\rightarrow V.$ The
proper length of such a curve is,
\begin{equation}
\ell (C)\equiv \int_{t_{1}}^{+\infty }\left( N^{2}-g_{ij}\left( \frac{dC^{i}%
}{dt}+\beta ^{i}\right) \left( \frac{dC^{j}}{dt}+\beta ^{j}\right) \right)
^{1/2}dt,
\end{equation}
and is therefore infinite if the lengths of the tangent vectors $dC/dt$ at
each point of $C$ are bounded away from zero by a positive constant. The
tangent vector to a curve depends on its parametrization. The use of $t$ as
a parameter on $C$ is linked to the $n+1$ splitting. This motivates the
following definition.

\begin{definition}
We say that the curve $C$ is \emph{uniformly timelike relatively to the $n+1$
splitting} if there exists a number $k>0$ such that on $C$,
\begin{equation}
N^{2}-g_{ij}\left(\frac{dC^{i}}{dt}+\beta ^{i}\right)\left(\frac{dC^{j}}{dt}
+\beta ^{j}\right)\geq k^{2}.
\end{equation}
\end{definition}

We have obviously the following theorem.

\begin{theorem}
The curves which are uniformly timelike relatively to the $n+1$ splitting
have an infinite proper length.
\end{theorem}

Examples of such curves are the orthogonal trajectories of the space
sections $M_{t}.$

The question of timelike geodesic completeness is more delicate since in an
arbitrary spacetime these curves do not necessarily possess the uniformity
mentioned above.

The tangent vector $u$ to a geodesic parametrized by arc length, or by the
canonical parameter in the case of a null geodesic, with components $%
dx^{\alpha}/ds$ in the natural frame, satisfies in an arbitrary frame the
differential equations,
\begin{equation}
u^{\alpha }\nabla _{\alpha }u^{\beta }\equiv u^{\alpha }\partial _{\alpha
}u^{\beta }+\omega _{\alpha \gamma }^{\beta }u^{\alpha }u^{\gamma }=0.
\label{geo-eqn}
\end{equation}
In the adapted frame the components of $u$ become,
\begin{equation}
u^{0}=\frac{dt}{ds},\quad u^{i}=\frac{dx^{i}}{ds}+\beta ^{i}\frac{dt}{ds},
\end{equation}
while the Pfaff derivatives are given by,
\begin{equation}
\partial_{0}\equiv\partial_{t}-\beta^{i}\partial_{i}, \quad\partial_{i}\equiv%
\frac{\partial}{\partial x^{i}}.
\end{equation}
It holds therefore that,
\begin{equation}
u^{\alpha}\partial_{\alpha }u^{\beta }\equiv \frac{dt}{ds}
\left(\partial_{t}-\beta^{i}\partial_{i}\right)u^{\beta} +\left(\frac{dx^{i}
}{ds}+\beta^{i}\frac{dt}{ds}\right)\partial_{i} u^{\beta}\equiv\frac{%
du^{\beta }}{ds}.
\end{equation}
Since $u^{0}\equiv dt/ds$, Eq. (\ref{geo-eqn}) with $\beta =0$ can be
written in the form,
\begin{equation}
\frac{d}{dt}\left(\frac{dt}{ds}\right)+\frac{dt}{ds}\left(\omega
_{00}^{0}+2\omega _{0i}^{0}v^{i}+\omega _{ij}^{0}v^{i}v^{j}\right)=0,
\label{3.7}
\end{equation}
where we have set,
\begin{equation}
v^{i}=\frac{dx^{i}}{dt}+\beta ^{i}.
\end{equation}
If $u$ is causal it holds that,
\begin{equation}
g_{ij}v^{i}v^{j}\leq N^{2}\leq N_{M}^{2},  \label{bounded length}
\end{equation}
and setting,
\begin{equation}
\frac{dt}{ds}=y,
\end{equation}
equation (\ref{3.7}) becomes,
\begin{equation}
\frac{y^{\prime}}{y}=-\left(\omega_{00}^{0}+2\omega_{0i}^{0}v^{i}+\omega
_{ij}^{0}v^{i}v^{j}\right) .  \label{3.11}
\end{equation}
The length (or canonical parameter extension) of the curve $C$ is
\begin{equation}\label{3.12}
\int_{t_{1}}^{+\infty }\frac{ds}{dt}\,dt
\end{equation}
and will be infinite if $ds/dt$ is bounded away from zero i.e., if $y\equiv
dt/ds$ is uniformly bounded.

We deduce from (\ref{3.11}) that,
\begin{equation}
\log\frac{y(t)}{y(t_{1})}=-\int_{t_{1}}^{t} \left(\omega_{00}^{0}+2\omega
_{0i}^{0}v^{i}+ \omega_{ij}^{0}v^{i}v^{j}\right) dt  \label{3.13}
\end{equation}
The integrand on the right hand side is itself a function of $t$ which
depends on the integration of the geodesic equations. However, we can
formulate \emph{sufficient} conditions on the spacetime metric under which $%
y $ is uniformly bounded using inequality (\ref{bounded length}). We denote
by $\nabla N$ the space gradient of the lapse $N$, by $K_{ij}=-N%
\Gamma_{ij}^{0}$ the extrinsic curvature of $M_{t}$ and by $\tau\equiv
g^{ij}K_{ij}$ the mean extrinsic curvature, negative in an expanding
universe. We then have the following result on the future completeness of
the spacetime $(\mathcal{V},g) $.

\begin{theorem}
Sufficient conditions for future timelike and null geodesic completeness of
the metric (\ref{2.1}) satisfying assumptions (\ref{h1}) and (\ref{h2}) are
that, for each finite $t_{1},$

\begin{enumerate}
\item  $|\nabla N|_{g}$ is bounded by a function of $t$ which is integrable
on $[t_{1},+\infty )$

\item  $|K|_{g}$ is bounded by a function of $t$ which is integrable on $
[t_{1},+\infty ).$
\end{enumerate}
\end{theorem}

\begin{proof}
Using the expressions of the connection coefficients, Eq. (\ref{3.13})
reads,
\begin{equation}
\log \frac{y(t)}{y(t_{1})}=\int_{t_{1}}^{t}N^{-1}\left(-\partial
_{0}N-2\partial _{i}Nv^{i}+K_{ij}v^{i}v^{j}\right) dt.
\end{equation}
On the other hand, we have on the curve $C$:
\begin{equation}
\partial _{0}N=\frac{dN}{dt}-v^{i}\partial _{i}N,
\end{equation}
which, using the inequality $g_{ij}v^{i}v^{j}=|v|_{g}^{2}\leq N^{2},$
implies that,
\begin{equation}
\log\frac{y(t)}{y(t_{1})}\leq 2\log
N_{m}^{-1}+N_{m}^{-1}\int_{t_{1}}^{t}\left(|\nabla
N|_{g}N_{M}+|K|_{g}N_{M}^{2}\right) dt,
\end{equation}
from which the result follows.
\end{proof}

\begin{corollary}
Let $P$ be the traceless part of $K.$ Then in an expanding universe,
Condition $2$ can be replaced by,

\begin{itemize}
\item  2a.\; The norm $|P|_{g}$ is integrable on $[t_{1},\infty )$.
\end{itemize}
\end{corollary}

\begin{proof}
If we set
\begin{equation}
K_{ij}=P_{ij}+\frac{1}{n}\tau g_{ij},
\end{equation}
we find that if the mean extrinsic curvature is negative, $\tau \leq 0$,
\begin{equation}
K_{ij}v^{i}v^{j}\equiv P_{ij}v^{i}v^{j}+\frac{1}{n}\tau g_{ij}v^{i}v^{j}\leq
P_{ij}v^{i}v^{j},
\end{equation}
and this completes the proof.
\end{proof}

If we could find \emph{necessary} conditions for the integral
(\ref{3.12}) to be bounded, then we will have proved a generic
singularity theorem, that is causal geodesic incompleteness.

\section{Example}

The Lorentzian metric found in \cite{cb-mo01} on the 3-manifold quotient of
the spacetime by an $S^{1}$ spacelike isometry group is of the type defined
in Section 2 with $\tau <0,$
\begin{equation}
g_{ij}=e^{2\lambda }\sigma _{ij}
\end{equation}
where $\sigma _{t}$ is uniformly equivalent to $\sigma _{t_{0}},$ for $%
t_{0}\geq 0$, and
\begin{equation}
e^{2\lambda }\geq 2t^{2},\quad 0<N_{m}\leq N\leq 2.
\end{equation}
The inequalities obtained in the quoted article show that Conditions 1 and
2a\, above are satisfied. The considered manifold is therefore future
timelike and null geodesically complete.


\begin{thebibliography}{9}
\bibitem{ch-kl93}  D. Christodoulou and S. Klainerman, \emph{The Global
Nonlinear Stability of the Minkowski Space}, (Princeton University Press,
1993).

\bibitem{cb-mo01}  Y. Choquet-Bruhat and V. Moncrief, \emph{Future global in
time einsteinian spacetimes with $U(1)$ isometry group}, Annales Henri
Poincar\'{e} (to appear).

\bibitem{ge70}  R. Geroch, J.Math.Phys. \textbf{11} (1970) 437-449

\bibitem{le52}  J. Leray, \emph{Hyperbolic differential equations}, (IAS,
Princeton, 1952).
\end{thebibliography}
\end{document}